\documentclass{article}
\usepackage{authblk}
\usepackage{graphicx} 
\usepackage[most]{tcolorbox}
\usepackage[numbers,sort&compress]{natbib}
\usepackage{xcolor}
\usepackage[margin=1.25in]{geometry}

\newtcolorbox{parabox}[1]{
    colback=gray!5,     %
    colframe=black!60!blue,
    fonttitle=\bfseries,
    title={#1},
    boxrule=1pt,
    left=10pt, right=10pt, top=10pt, bottom=10pt,
    float
}

\title{Data augmentation as a framework for modeling \\hippocampal contributions to generalization}
\author[1*]{Tyler Bonnen}
\author[2*]{Andrew Kyle Lampinen}
\affil[1]{University of Pennsylvania}
\affil[2]{Anthropic}
\affil[*]{Equal contribution}
\date{}

\begin{document}

\maketitle
\begin{abstract}
\vspace{.1em}
\noindent 
The hippocampus plays a critical role in generalization, enabling us to flexibly repurpose prior experiences to perform novel tasks. Here we suggest that data augmentation---a machine learning strategy to improve generalization by refactoring prior experience---offers a useful framework to conceptualize and model hippocampal function. We begin by outlining how data augmentation operates across two timescales: the traditional ``offline'' setting, where refactoring training data yields more general representations, and an ``online'' setting, where retrieved experiences can be flexibly refactored at test time to support zero-shot inference. We suggest that these `offline' and `online' computational strategies map onto functions supported by the hippocampus. Critically, we argue that these computational tools can be leveraged to develop formal `linking functions' between experimental evidence and theoretical claims,  such that a unified modeling approach can be used to predict the diverse behaviors that depend on the hippocampus---from navigating in high-dimensional sensory environments to more abstract inferences. We hope this perspective, and the modeling strategies it makes available, will support new efforts to formalize and evaluate theories of hippocampal function.
\end{abstract}

\begin{figure}[tph]
    \centering
    \includegraphics[width=\textwidth]{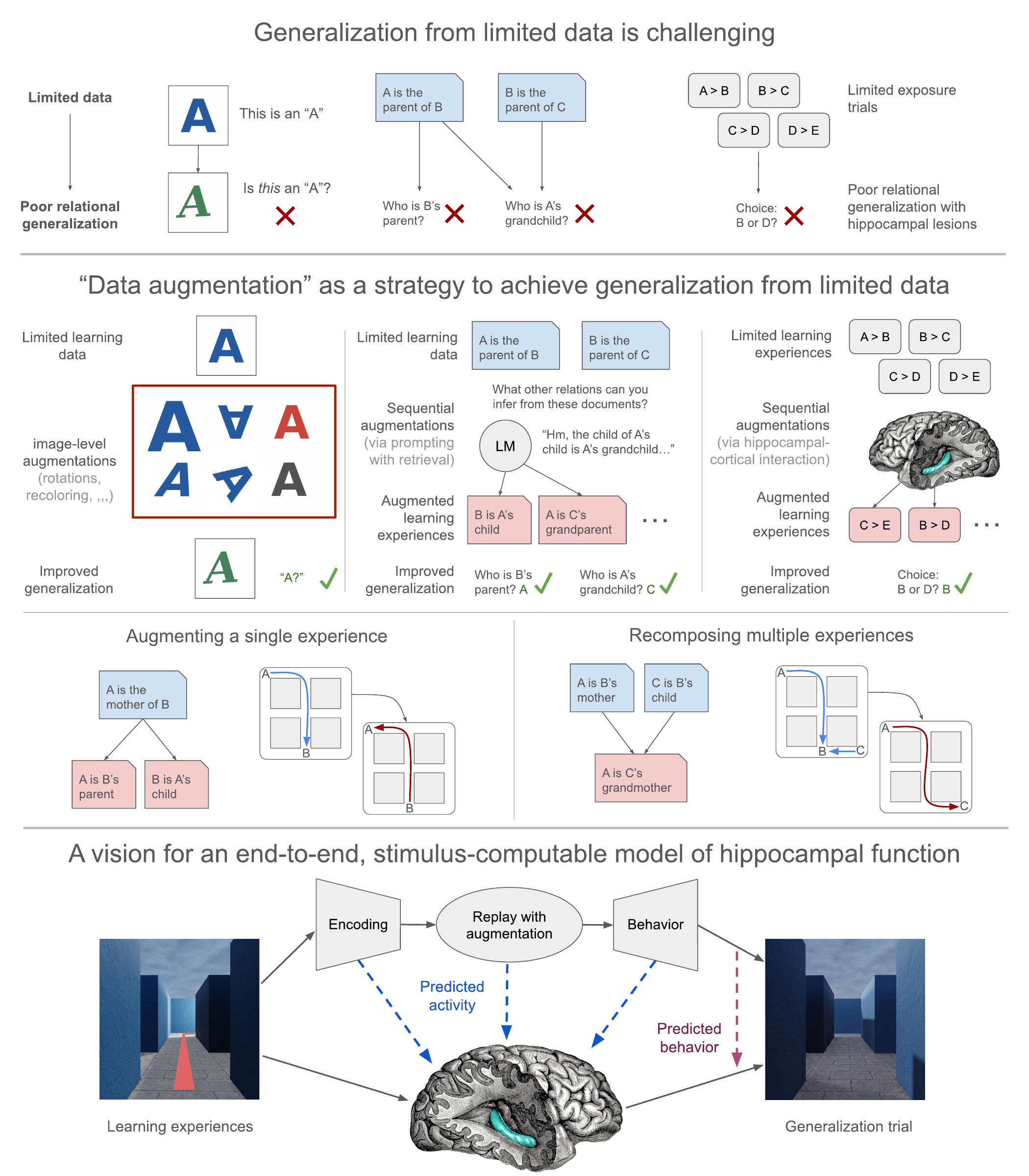}
    \vskip-2pt
    \caption{An overview of our perspective, relating  machine learning findings and methods to phenomena of hippocampal contributions to generalization. (top) The deficit: machine learning systems without data augmentation, and animals with hippocampal-lesions, show impaired generalization. (middle) Data-augmentation has been shown to ameliorate these deficits in machine learning, and we argue that the hippocampus may contribute to a similar process. (middle, second row) Data augmentation can take various forms, including refactoring an individual learning experience based on prior knowledge, or recombining multiple experiences. (bottom) Ultimately, we aim for an end-to-end, stimulus computable model that predicts how neural activity and behavior depend on individual learning experiences.} \label{fig:overview}
\end{figure}

\vspace{1em}

\noindent Intelligent systems must generalize from limited experience. Humans and other animals have remarkable strategies for achieving this feat. After walking around a new environment, for example,  we can perform  many navigation-related behaviors---from retracing our steps to identifying shortcuts, or planning novel routes between two locations. That is, a single experience can be used to support many downstream behaviors. This process depends on the interplay between two related abilities: the ability to access prior experiences, and the ability to transform those experience in a way that meets subsequent task demands. A wealth of evidence implicates the hippocampus in these operations, from lesion studies showing impairments in spatial navigation \citep{eichenbaum2000hippocampus} or linguistic generalizations \citep{bayley2002medial}, to neural signatures including replay \citep{gupta2010hippocampal}, preplay \citep{olafsdottir2015hippocampal}, place cells \citep{okeefe1971hippocampus}, and time cells \citep{eichenbaum2014time}. %
What computational principles allow a limited set of experiences to support such diverse behaviors?

\vspace{.5em} \noindent Many conceptual frameworks have been proposed to account for these empirical findings (Box 1), including cognitive maps \citep{tolman1948cognitive, okeefe1978hippocampus}, pattern separation \citep{marr1971simple, treves1994computational}, Complementary Learning Systems \citep[CLS][]{mcclelland1995complementary}, predictive representations \citep{stachenfeld2017hippocampus,momennejad2017successor,tarder2024brain}, inferential reasoning \citep{zeithamova2012hippocampus}, and compositionality \citep{whittington2020tolman, behrens2018cognitive, solomon2020semantic}. Each framework captures important aspects of hippocampal function, but they offer distinct accounts of how a single experience can be repurposed to support such diverse behaviors. Understanding the relationship between each of these theories and experimental evidence requires some kind of `linking function' (i.e., a means to infer the theoretical implications of empirical observations), yet this remains an informal process, even for prevailing computational frameworks (Box 2). `Stimulus-computable' models from deep learning provide one route to concretize the relationship between theory and data; theories of hippocampal function can be used to design models that operate over the same stimuli presented to experimental participants (e.g., sensory environments, distances to travel), and can be evaluated via their quantitative fit to experimental measurements (e.g., trial-by-trial behavioral dynamics, neural activity). As such, the models themselves can provide the linking function that relates theoretical claims to empirical observations, and distinct theoretical frameworks can be evaluated using the some empirical framework. However, this requires computational techniques that are well-suited to instantiate theories of the hippocampus.

\vspace{.5em} \noindent Here, we suggest that technical frameworks from machine learning offer novel routes to formalize and evaluate theories of hippocampal function. We develop this idea by reviewing machine learning literature on `data augmentation' \citep{shorten2019survey}, from vision to language. We identify how data augmentation can operate at two complementary timescales: offline during training to build robust representations, and online via retrieval to support flexible inference on the present task. We relate these modeling strategies to the diverse behaviors that depend on the hippocampus. Our goal is not to propose a novel computational theory of the hippocampus, per se, but to suggest an opportunity for a modeling framework that could allow bridge between computational models and experimental or real-world settings, thus allowing better evaluation of theories of hippocampal function.

\begin{parabox}{Box 1 | Relating to prior theories of hippocampal function}
Many computational perspectives have been offered on hippocampal contributions to learning and generalization. Complementary Learning Systems \citep[CLS;][]{mcclelland1995complementary} offers the view that the hippocampus primarily complements cortical learning by allowing rapid learning without interference. However, it initially had relatively little focus on contributions to generalization. More recent models have considered how hippocampal(-cortical) recurrence could support generalization \citep{kumaran2012generalization} by connecting related memory episodes. A similar perspective is taken more recently by models describing a broad range of replay findings as context-driven memory reactivation \citep{zhou2026unifying}. Other accounts have focused on the hippocampus forming predictive maps \citep{stachenfeld2017hippocampus,momennejad2017successor} or more generally predictive representations \citep{tarder2024brain}. Perhaps the most closely related perspective is from \citet{zeithamova2012hippocampus}, who argue that recall of past events can allow making connections with present ones, either when acting or during encoding. %
While our perspective is generally compatible with these prior models, it offers the unifying \emph{normative} interpretation of hippocampal-cortical interactions as \emph{data augmentation}---recombining memories with each other, and with prior knowledge, to improve current or future generalization.
We believe that this perspective offers an overarching computational description of the \emph{process} by which, for example, the hippocampus might help to ``organize knowledge for flexible behavior'' \citep{behrens2018cognitive}. This perspective is consistent with the idea that it is useful for the hippocampus to be less compressive than cortex, and ``[preserve] experiences in a relatively raw format for later interpretation'' \citep{nagy2025adaptive}. However, the data augmentation perspective allows drawing on predictions about the detail of how individual experiences are processed to support different kinds of generalization.
\end{parabox}

\section{Data Augmentation in Machine Learning}

Machine learning systems face a fundamental challenge: performance improves with extensive training data, yet these data are expensive to obtain. More fundamentally, models are typically deployed in domains not well represented in the training data. Data augmentation is a strategy to address these limitations by synthesizing new training examples from existing data. The core idea is simple: one data point, when appropriately transformed, generates multiple learning signals. This approach has proven effective across many domains, from computer vision to language modeling. Below we outline how these ideas, operating at different timescales, serve complementary functional roles.

\subsection{Offline augmentation: \\Enhancing data to learn robust representations}

\vspace{.5em} \noindent We begin with vision, where data augmentation first became prevalent. How can a visual system learn reliable representations from a limited training dataset? For example, any image of a cat is just one of infinite possible images of that cat; the viewpoint, background, or camera could all change, without changing the cat's identity. Data augmentation attempts to discover the structure underlying an image, by learning a representation that is consistent across different re-projections of that image. For example, rotating or recoloring an image will change its pixels, but not the identity or the spatial relationships of the entities within the image. By learning to ignore these nuisance variables, vision models are able to better learn about the entities represented in the images \citep{shorten2019survey}---thus producing vision encoders that can extract useful features for many downstream visual tasks---from classification to object segmentation \citep{oquab2024dinov2}.

\vspace{.5em} \noindent Data augmentation can also be applied across higher levels of visual abstraction. Similar strategies can be used for `multi-view' models that infer the underlying spatial structure of an environment \cite{wang2025vggt}. These data augmentation strategies are not limited purely to the domain of vision; vision language models, like CLIP, commonly used data augmentation strategies over linguistic information; the same image can be paired with multiple captions (e.g., ``a cat sitting on a bed,'' ``a house cat in a bedroom'') \citep{radford2021learning}. This cross-modal augmentation enables using language as a flexible interface for retrieving visual knowledge.

\vspace{.5em} \noindent The goal of these data augmentations is to learn abstractions that can be used to support many downstream behaviors. Single-image transformations yield appearance-level invariances; multi-view integration across modalities yields geometric structure; vision-language transformations learn stable image-semantic mappings. While this approach is not motivated by biological plausibility, nonetheless, many systems trained with these data augmentation strategies exhibit emergent alignment\footnote{In the sense of similar structure in their representations \citep[cf.][]{sucholutsky2025getting}.} with human visual processing; models trained via image-text alignment predict human perception of semantics \citep{muttenthaler2023human}, while multi-view transformers predict human 3D shape perception  \citep{bonnen2024multiview}. 

\vspace{.5em} \noindent \textbf{Sequential tasks:} Sequential tasks, such as language processing, only multiply the problems of limited data. Any path through a city, or sentence on a topic, is just one of infinite ways that the underlying structure could be experienced. Thus, in sequential tasks from Reinforcement Learning (RL) to language modeling, machine learning researchers have likewise relied on data augmentation. 

\vspace{.5em} \noindent As a simple example, consider the statement ``A is the parent of B.'' This relationship could be encoded in many other surface forms, e.g., the reversal ``B is the child of A.'' Surprisingly, language models that have been trained on the relation in one direction do not generalize to the reverse \citep{berglund2024reversal}. Thus, researchers have used data augmentation to help the system learn the more abstract information; explicitly training models on reversed sequences \citep{golovneva2024reverse}, or augmenting sequences with masked copies that encourage learning both dependencies \citep{pan2025closing}. 

\vspace{.5em} \noindent However, sequential data affords much richer augmentations, beyond simple, fixed transformations. Indeed, language models themselves are powerful and flexible tools for transforming information. Thus many works \citep{lampinen2025generalization,akyurek2024deductive,yang2025synthetic,park2025new,kim2026data} use the model's own capabilities to elaborate training documents by reasoning about possible extrapolations and connections between them---by extracting entities and reasoning about their relations \citep{yang2025synthetic}, or chaining statements from multiple documents to make new inferences \citep{akyurek2024deductive,lampinen2025generalization}.
These strategies bootstrap from the model's existing language abilities to enrich the experiences at hand. By generating alternative ``views'' of the information underlying the learning experiences, the system is able to learn more effectively.

\begin{parabox}{Box 2|Linking functions: tightly coupling theory and evidence} The relationship between experimental evidence and scientific theory is underdetermined \citep{duhem1954aim}; the same data can be plausibly interpreted in many ways. As such, the implications of empirical data must be inferred. This mapping between measurements and their theoretical implications is commonly described as a `linking function,' and there is a rich tradition that aims to formalize this process \citep{brindley1960physiology}. While computational models are thought to address this ambiguity \citep[e.g.][]{guest2021computational}, prevailing models of hippocampal function remain imprecise because the models operate over idealized, hand-tailored inputs (e.g., one-hot vectors) which are designed to reflect known stimulus properties. Consequently, the relationship between model operations and experimental variables must still be inferred, and the assumptions embedded in the input representation (e.g., which stimulus properties matter, how they relate to one another) do considerable theoretical work. In contrast, `stimulus-computable' models greatly reduce this ambiguity by operating over the same sensory data presented to experimental subjects \citep{yamins2016using}. In this formulation, the model performs the same task that animals are confronted with, from the same sensory data, such that the linking function is the model itself, and model predictions embody theoretical claims. This approach has already been used to understand neuroanatomical structures within the medial temporal lobe. It was long debated whether medial temporal structures like perirhinal cortex (PRC) support performance on perceptual tasks; there were competing interpretations of the available data. Using a stimulus-computable modeling framework to formalize the relationship between experimental data and theoretical claims reduced this ambiguity, resolving this debate \citep{bonnen2021ventral}. While the hippocampus invites the same approach, this demands a new class of computational models that can perform memory-related tasks, given the same sensory data presented to experimental subjects. In such a model, different strategies for data augmentation could serve as computational specifications for the process of reprojecting data to enhance generalization; the models' internal representations and behavior under those different strategies would then make distinct, testable predictions about the neural activity and behavior of the animal on the task being modeled. 
\end{parabox}

\subsection{Online augmentation: \\Flexible inference through context-dependent retrieval}

\vspace{1em} \noindent How can a system flexibly re-purpose stored knowledge for tasks it was not originally trained to solve? In the offline approaches above, abstractions that may be useful in the future are encoded into the system by augmenting the data accordingly during training. However, at test time the system might face entirely novel challenges that were not anticipated during training. Language models provide a useful setting to understand how a system can adaptively re-purpose its knowledge to solve unexpected tasks via online data augmentation.

\vspace{1em} \noindent Online data augmentation addresses this limitation by flexibly modifying how knowledge is queried. In short, the system retrieves\footnote{Interpreting retrieval as online data augmentation is atypical, but we find it useful for emphasizing their similar role: recombining experiences to support generalization.} relevant information and provides it as context, augmenting the present information with past experiences, and relying on  the model's ability to appropriately combine these multiple sources. For the parent-child example, this means retrieving the learned fact (``B is the parent of A'') and presenting it in context, rather than expecting the model to retrieve and transform it internally. When information is provided in context, language models successfully perform reversals, make transitive inferences, and better answer novel questions \citep{lampinen2025generalization,park2025new,lampinen2025latent}. (Alternatively, models can learn to do approximate retrieval via generation \citep{chaudhry2026improving}.) This strategy extends beyond language to visual reasoning, where vision-language models can generate intermediate visual representations at test time to support spatial inferences \citep{qin2025covt, bigverdi2024perception}. 

\vspace{1em} \noindent Unlike offline data augmentation, where transformations are fixed during training, online augmentation is flexible and task-dependent. The key insight is that systems can bootstrap from their existing capabilities to transform retrieved content to serve novel goals---such as reasoning about retrieved documents.  %
What makes this ``augmentation'' is that the same knowledge is re-factored differently (reversed, re-composed, or chained), depending on current goals. This ability to recombine learned elements in novel ways emerges naturally in systems that can retrieve knowledge in a form that preserves its relational structure, rather than compressing everything into fixed associations.

\subsection{Data augmentation at different timescales}

\noindent Offline and online data augmentation provide complementary strategies for sample-efficiency. Offline augmentations apply predetermined transformations to learn features useful for downstream behaviors; online augmentations apply context-dependent transformations to enable flexible inference at test time. These computational strategies are unified in recognizing that any given experience contains a single projection of the underlying data; a photograph of a scene, a sentence about an idea. Re-projecting these data from different perspectives can extracts more information than any single projection could provide. The difference between online and offline augmentation is simply when that re-projection occurs.

\section{The analogy between data augmentation and hippocampal function}

There are useful parallels between hippocampal function and the data augmentation strategies outlined above. In essence, data augmentation requires two components: an experience, and operations to transform it. We suggest the hippocampus preserves information related to single events in a way that supports subsequent transformation \citep[cf.][]{nagy2025adaptive}. Offline, these experiences can be reinstated and interleaved with ongoing learning, allowing cortical structures to build more general representations from sets of related events \citep{mcclelland1995complementary}. Online, prior experiences can be re-instated, and then re-factored, in order to meet the demands of the current task \citep{aly2018flexible}. In both cases the hippocampus provides information that can be operated on by---and through the interaction with---other neural structures. We elaborate on the offline and online settings below (and in Box 3).

\begin{parabox}{Box 3 | Interpreting other prior findings as data augmentation}
Many other prior findings on hippocampal activity can be reinterpreted from the perspective of data augmentation. We outline a few more of these connections here.\\ 

\textbf{Abstractions and schemas:} Hippocampal replay recomposing learning experiences according to previously-learned abstract structures \citep{liu2019human} is similar to how offline augmentation in language models can reproject learning experiences through prior knowledge to yield new data. 
Other structural knowledge like schemas \citep{masis2022schema} may similarly support this type of augmentation; in turn, replay that integrates related experiences may help to extract more abstract schemas from those experiences \citep{lewis2011overlapping}. \\

\textbf{Compositional reasoning:} Online hippocampal replay has also been implicated in reasoning about the composition of a structure by generating possible compositions of previously learned components \citep{schwartenbeck2023generative,he2026human}; this can be interpreted as augmentation through recomposing prior experiences to make new predictions. \\

\textbf{Semantic search:} Findings that the hippocampus supports online enumeration of semantic properties of a concept \citep{cutler2019searching}---and in particular, the argument that it does so by offering shortcuts between dissimilar semantic features through bound representations \citep{solomon2020semantic}---can also be interpreted from the (online) data-augmentation perspective as augmenting each concept with direct connections to its experienced features. %
\end{parabox}

\vspace{.5em} \noindent \textbf{Offline augmentation and systems consolidation.} 
A long tradition holds that the hippocampus supports the gradual transfer of knowledge to neocortical structures via systems consolidation---for example, complementary learning systems theory \citep[CLS][]{mcclelland1995complementary, kumaran2016learning}. A candidate mechanism facilitating this division of labor is replay: during rest, the hippocampus `re-instates' patterns of neural activity that occurred during the initial encoding of those experiences \citep{wilson1994reactivation}.%
Remarkably, replay does not always recapitulate exact experiences; it also produces never-experienced trajectories \citep{gupta2010hippocampal,olafsdottir2015hippocampal}, reverse unrolls from a reward %
\citealp{foster2006reverse,liu2019human}, and sometimes exhibits a bias away from recently experienced trajectories \citealp{gupta2010hippocampal}. %
While the nature of these hippocampal-neo-cortical interactions is an ongoing research question \citep{zhou2026unifying}, it is clear that manipulating which memories are reactivated during sleep changes subsequent performance \citep{rasch2007odor, rudoy2009strengthening}. In effect, replay enables a limited set of experiences to be re-factored, such that the neocortex can learn structure that was only latent in the individual learning events. This is data augmentation in the precise sense that we began with: the hippocampus supplies an experience, via replay, which can be operated on by other neural structures. This is not a novel claim about hippocampal function; rather, data augmentation provides a coherent framework to \emph{model} these diverse properties of the hippocampus, and the downstream behaviors they enable. With these tools, theoretical claims can be made into explicit design choices: a claim that replay prioritizes weakly-learned experiences, for instance, is instantiated as a sampling policy over stored experiences, whose consequences for consolidation and subsequent behavior can be directly evaluated (Box 4). 

\vspace{.5em} \noindent \textbf{Online data augmentation in spatial navigation.} An animal traversing a novel environment processes sensory information associated with that environment's unique configuration of landmarks, distances, objects, and actions. This single trajectory can only provide a limited representation of the environment. But that structure is latent within the sequence, and it can be discovered. Just as re-projecting a single image aims to recover the scene it captured, re-projecting a trajectory can reveal the environment's spatial structure. While many dimensions could be used to reproject these data, egocentric and allocentric coordinate frames are particularly useful for navigation. An egocentric encoding (`turn left, walk ten steps') captures action sequences but not environmental structure; an allocentric encoding (`food north of the tree') captures world coordinates but not the affordances necessary to traverse a scene. Re-projecting between these coordinate frames, combined with geometric operations like reversal and composition, unlocks what a single sequence cannot: novel shortcuts, backward routes, and answers to ``where was that object?'' The hippocampus supports exactly these operations. During an episode, replay and preplay transform stored experiences using these geometric operations---identifying shortcuts \citep{gupta2010hippocampal}, connecting separate sequences once a link has been established \citep{tarder2026hippocampus}, reversing experiences \citep{foster2006reverse,foster2017replay}, and even preplaying routes to potential future goals \citep{olafsdottir2015hippocampal}. Online, these transformations can both enablie the kinds of online data-augmentation strategies outlined above to support current behavioral goals \citep[cf.][]{aly2018flexible}; offline, they can support consolidating the spatial primitives that online planning and behavior require.

\vspace{.5em} \noindent \textbf{Online data augmentation in relational generalizations.} The principles outlined above extend beyond spatial domains to abstract relational reasoning. When an animal learns a single fact---for example, ``A is larger than B''---this provides just one projection of the underlying relational structure. But that latent structure can be re-projected (reversed to ``B is smaller than A,'' or chained with ``B is larger than C'' to infer ``A is larger than C,''). Language models that encode such facts solely into their parameters fail these transformations---they cannot reverse learned relations or make multi-step transitive inferences \citep{berglund2024reversal, balesni2024two}. The same knowledge maintained in retrievable form, however, supports systematic transformation \citep{lampinen2025generalization, akyurek2024deductive}. Hippocampal lesions produce analogous deficits: failures in transitive inference requiring integration across experiences \citep{eichenbaum2009neurobiology} and inflexibility in rephrasing relations \citep{bayley2002medial}. Intact hippocampal function, by contrast, enables flexible relational generalization through replay-driven recombination of prior experiences \citep{kurth2023replay}---integrating experiences to support transitive inferences \citep{shohamy2008integrating,kumaran2012generalization,schlichting2015learning}, or generalization after contingency changes \citep{renz2025neural}. Just as spatial navigation requires re-projection across coordinate frames, relational reasoning requires re-projection across relational structures---operations the hippocampus supports through episodic retrieval.

\begin{parabox}{Box 4 | Modeling hippocampal function; a worked example}

Our proposal's value is ultimately in the modeling approaches it enables. To make this strategy concrete, we engage with a question that has been the topic of considerable attention: \textit{which experiences are replayed by the hippocampus?} Normative accounts have proposed that replay is a function of gain and need \citep{mattar2018prioritized}; how much reactivating a memory improves subsequent decisions, and how likely that memory is to be relevant. Framed in relation to our data augmentation proposal, this is a policy for curating the augmentation set, determining which experiences to re-factor. However, determining which experiences are replayed does not specify \textit{how} information must be replayed to support subsequent behavior. \\

Consider navigation. First, an animal must process their sensory trajectories through an environment, and encode this information within the hippocampus in a manner that preserves it for future replay. Offline, these experiences are thought to be replayed proportional to gain and need, but there are many possibilities for how these experiences might be re-factored to support adaptive behaviors: possibilities include re-projecting sensory information between coordinate frames to plan a route, reversing a sequence to propagate rewards backward \citep{foster2006reverse}, or composing segments of trajectories to construct a shortcut \citep{gupta2010hippocampal}. Prior modeling work instantiates these as relatively abstract operations over idealized inputs that have no concrete relationship to the sensory experiences that animals are confronted with. As such, there is not a formal `linking function' between the model operations and experimental variables, and so the relation to theoretical claims is speculative. \\

In contrast, the computational frameworks we have introduced can operate over the same sensory data as experimental subjects. As such, model design choices---and by proxy, the theoretical claims they instantiate---can be evaluated empirically, via the model's quantitative fit to animal behavior on the same experimental task. This approach forces a commitment that verbal theories can leave implicit: which operations are supplied by the hippocampus, and which are accomplished through interaction with other structures. Whether a shortcut reflects composition within the hippocampus or emerges from hippocampal–cortical exchange; whether the hippocampus re-instates experiences veridically or transforms them (e.g., by compressing them into more generalizable structure \citep{stachenfeld2017hippocampus}) are questions a stimulus-computable model must answer explicitly. Framing these theoretical claims as modeling choices allows them to be resolved empirically. 
\end{parabox}

\section{Conclusions}

We have outlined how data augmentation strategies from machine learning could be used to understand and model hippocampal function. Our goal is not to introduce a novel theory of hippocampal function, but to outline a perspective that could help to integrate many existing theories by interpreting replay, consolidation, and flexibly task-dependent retrieval as instances of a common set of operations. %
Because data augmentation is a computational strategy coupled to stimulus-computable methods, it would allow us instantiate theories of hippocampal function as models that operate over the sensory data animals actually receive , rather than the hand-designed inputs prevailing models require. Realizing this promise is now potentially a feasible undertaking. A natural first step is to build such a model for a single, well-characterized behavior (e.g., the identifying a novel shortcut) where competing theories make concrete predictions. These disagreements can then be adjudicated by their fit to the behavior, and neural activity, of animals performing the same task. This will demand experimental paradigms rich enough to allow nontrivial generalization from naturalistic input, which remain scarce \citep{carvalho2025naturalistic}. But we see much of the value of this framing in the demands it imposes: it forces distinctions that verbal theories can leave implicit \citep[cf.][]{guest2021computational}---which operations the hippocampus supplies, and which emerge through its interaction with other structures---into explicit modeling choices that can be weighed against data. We hope that this will prove a productive step toward a more detailed understanding of the computational role of the hippocampus.

\section*{Acknowledgements}

We thank Anna Schapiro, Stefano Palmintieri, and the anonymous reviewers for feedback. This work is supported by the National Institute of Neurological Disorders and Stroke within the NIH (Award Number F99NS125816) and the UC Presidential Postdoctoral Fellowship Award.

\bibliographystyle{unsrtnat}
\bibliography{main}

@article{berglund2024reversal,
  title={{The Reversal Curse: LLMs trained on “A is B” fail to learn “B is A”}},
  author={Berglund, Lukas and Tong, Meg and Kaufmann, Maximilian and Balesni, Mikita and Stickland, Asa Cooper and Korbak, Tomasz and Evans, Owain},
  year={2024},
  journal={International Conference on Learning Representations},
  note={\\[0.5em] *This paper demonstrates the ``Reversal Curse''---the failure of language models to generalize to relations in a reverse order from how they were encountered in training.}
}

@article{balesni2024two,
  title={{The Two-Hop Curse: LLMs trained on A $\rightarrow$ B, B $\rightarrow$ C fail to learn A $\rightarrow$ C}},
  author={Balesni, Mikita and Korbak, Tomek and Evans, Owain},
  journal={arXiv preprint, arXiv:2411.16353},
  year={2024},
  note={\\[0.5em] *This paper demonstrates a failure of language models to generalize by linking together multiple statements that appeared in their training data.}
}

@inproceedings{yang2025synthetic,
  title={Synthetic continued pretraining},
  author={Yang, Zitong and Band, Neil and Li, Shuangping and Candes, Emmanuel and Hashimoto, Tatsunori},
  year={2025},
  booktitle={The Thirteenth International Conference on Learning Representations},
  note={\\[0.5em] **This paper argues that language models struggle to learn facts about entities and their relations from limited data. Thus, it proposes an offline data augmentation strategy that infers underlying entities and relations, and then explicitly augments experiences to offer different presentations of that information.}
}

@article{lampinen2025generalization,
  title={On the generalization of language models from in-context learning and finetuning: a controlled study},
  author={Lampinen, Andrew K and Chaudhry, Arslan and Chan, Stephanie CY and Wild, Cody and Wan, Diane and Ku, Alex and Bornschein, J{\"o}rg and Pascanu, Razvan and Shanahan, Murray and McClelland, James L},
  journal={Foundations of Reasoning in Language Models Workshop, NeurIPS 2025},
  year={2025},
  note={\\[0.5em] **This paper shows that, while language models show failures of relational and multi-hop generalization from their training data, they can flexibly perform these inferences over information in context. Thus, the paper proposes using offline data augmentation to go beyond the single view afforded in the data.}
}

@article{lampinen2025latent,
  title={Latent learning: episodic memory complements parametric learning by enabling flexible reuse of experiences},
  author={Lampinen, Andrew Kyle and Engelcke, Martin and Li, Yuxuan and Chaudhry, Arslan and McClelland, James L},
  journal={Transactions on Machine Learning Research},
  year={2026},
  note={\\[0.5em] **This paper argues for an online retrieval strategy as a solution to flexibly solving unanticipated tasks at test time, including demonstrating its value on latent learning tasks.}
}

@article{chaudhry2026improving,
  title={Improving Latent Generalization Using Test-time Compute},
  author={Chaudhry, Arslan and Thiagarajan, Sridhar and Lampinen, Andrew},
  journal={3rd Conference on Language Modeling},
  year={2026},
  note={\\[0.5em] **This paper illustrates how RL can substitute for part of the benefit of retrieval in online augmentation; instead of explicitly retrieving information, the model can learn via RL how to regenerate the episode from memory, and only then reproject that episode to solve a new task.}
}

@article{park2025new,
  title={{New News: System-2 Fine-tuning for Robust Integration of New Knowledge}},
  author={Park, Core Francisco and Zhang, Zechen and Tanaka, Hidenori},
  journal={arXiv preprint arXiv:2505.01812},
  year={2025}
}

@article{bayley2002medial,
  title={Medial temporal lobe amnesia: Gradual acquisition of factual information by nondeclarative memory},
  author={Bayley, Peter J and Squire, Larry R},
  journal={Journal of Neuroscience},
  volume={22},
  number={13},
  pages={5741--5748},
  year={2002},
  publisher={Society for Neuroscience}
}

@article{eichenbaum2009neurobiology,
  title={The neurobiology of memory based predictions},
  author={Eichenbaum, Howard and Fortin, Norbert J},
  journal={Philosophical Transactions of the Royal Society B: Biological Sciences},
  volume={364},
  number={1521},
  pages={1183--1191},
  year={2009},
  publisher={The Royal Society London}
}

@article{liu2019human,
  title={Human replay spontaneously reorganizes experience},
  author={Liu, Yunzhe and Dolan, Raymond J and Kurth-Nelson, Zeb and Behrens, Timothy EJ},
  journal={Cell},
  volume={178},
  number={3},
  pages={640--652},
  year={2019},
  publisher={Elsevier}
}

@article{olafsdottir2015hippocampal,
  title={Hippocampal place cells construct reward related sequences through unexplored space},
  author={{\'O}lafsd{\'o}ttir, H Freyja and Barry, Caswell and Saleem, Aman B and Hassabis, Demis and Spiers, Hugo J},
  journal={Elife},
  volume={4},
  pages={e06063},
  year={2015},
  publisher={eLife Sciences Publications, Ltd}
}

@article{nagy2025adaptive,
  title={Adaptive compression as a unifying framework for episodic and semantic memory},
  author={Nagy, David G and Orb{\'a}n, Gerg{\H{o}} and Wu, Charley M},
  journal={Nature Reviews Psychology},
  pages={1--15},
  year={2025},
  publisher={Nature Publishing Group US New York}
}

@inproceedings{akyurek2024deductive,
  title={Deductive closure training of language models for coherence, accuracy, and updatability},
  author={Aky{\"u}rek, Afra Feyza and Aky{\"u}rek, Ekin and Choshen, Leshem and Wijaya, Derry Tanti and Andreas, Jacob},
  booktitle={Findings of the Association for Computational Linguistics: ACL 2024},
  pages={9802--9818},
  year={2024}
}

@inproceedings{golovneva2024reverse,
  title={Reverse Training to Nurse the Reversal Curse},
  author={Golovneva, Olga and Allen-Zhu, Zeyuan and Weston, Jason E and Sukhbaatar, Sainbayar},
  year={2024},
  booktitle={First Conference on Language Modeling}
}

@article{renz2025neural,
  title={Neural replay is connected to latent cause inference and supports fast generalization},
  author={Renz, Fabian M and Grossman, Shany and Daw, Nathaniel D and Dayan, Peter and Doeller, Christian F and Schuck, Nicolas W},
  journal={bioRxiv},
  pages={2025--12},
  year={2025},
  publisher={Cold Spring Harbor Laboratory}
}

@article{zhou2026unifying,
  title={A unifying account of replay as context-driven memory reactivation},
  author={Zhou, Zhenglong and Kahana, Michael J and Schapiro, Anna C},
  journal={eLife},
  volume={13},
  pages={RP99931},
  year={2026},
  publisher={eLife Sciences Publications, Ltd}
}

@article{gupta2010hippocampal,
  title={Hippocampal replay is not a simple function of experience},
  author={Gupta, Anoopum S and Van Der Meer, Matthijs AA and Touretzky, David S and Redish, A David},
  journal={Neuron},
  volume={65},
  number={5},
  pages={695--705},
  year={2010},
  publisher={Elsevier}
}

@article{kumaran2012generalization,
  title={Generalization through the recurrent interaction of episodic memories: a model of the hippocampal system.},
  author={Kumaran, Dharshan and McClelland, James L},
  journal={Psychological review},
  volume={119},
  number={3},
  pages={573},
  year={2012},
  publisher={American Psychological Association}
}

@article{tolman1948cognitive,
  title={Cognitive maps in rats and men},
  author={Tolman, Edward C},
  journal={Psychological Review},
  volume={55},
  number={4},
  pages={189--208},
  year={1948}
}

@article{zeithamova2012hippocampus,
  title={The hippocampus and inferential reasoning: Building memories to navigate future decisions},
  author={Zeithamova, Dagmar and Schlichting, Margaret L and Preston, Alison R},
  journal={Frontiers in Human Neuroscience},
  volume={6},
  pages={70},
  year={2012}
}

@article{wilson1994reactivation,
  title={Reactivation of hippocampal ensemble memories during sleep},
  author={Wilson, Matthew A and McNaughton, Bruce L},
  journal={Science},
  volume={265},
  number={5172},
  pages={676--679},
  year={1994}
}

@article{foster2017replay,
  title={Replay comes of age},
  author={Foster, David J},
  journal={Annual Review of Neuroscience},
  volume={40},
  pages={581--602},
  year={2017}
}

@article{okeefe1971hippocampus,
  title={The hippocampus as a spatial map: Preliminary evidence from unit activity in the freely-moving rat},
  author={O'Keefe, John and Dostrovsky, Jonathan},
  journal={Brain Research},
  volume={34},
  number={1},
  pages={171--175},
  year={1971}
}

@article{eichenbaum2014time,
  title={Time cells in the hippocampus: a new dimension for mapping memories},
  author={Eichenbaum, Howard},
  journal={Nature Reviews Neuroscience},
  volume={15},
  number={11},
  pages={732--744},
  year={2014}
}

@book{okeefe1978hippocampus,
  title={The hippocampus as a cognitive map},
  author={O'Keefe, John and Nadel, Lynn},
  year={1978},
  publisher={Oxford University Press}
}

@article{marr1971simple,
  title={Simple memory: a theory for archicortex},
  author={Marr, David},
  journal={Philosophical Transactions of the Royal Society of London B},
  volume={262},
  number={841},
  pages={23--81},
  year={1971}
}

@article{treves1994computational,
  title={Computational analysis of the role of the hippocampus in memory},
  author={Treves, Alessandro and Rolls, Edmund T},
  journal={Hippocampus},
  volume={4},
  number={3},
  pages={374--391},
  year={1994}
}

@article{mcclelland1995complementary,
  title={Why there are complementary learning systems in the hippocampus and neocortex: insights from the successes and failures of connectionist models of learning and memory},
  author={McClelland, James L and McNaughton, Bruce L and O'Reilly, Randall C},
  journal={Psychological Review},
  volume={102},
  number={3},
  pages={419--457},
  year={1995}
}

@article{whittington2020tolman,
  title={The Tolman-Eichenbaum machine: unifying space and relational memory through generalization in the hippocampal formation},
  author={Whittington, James CR and Muller, Timothy H and Mark, Shirley and Chen, Guifen and Barry, Caswell and Burgess, Neil and Behrens, Timothy EJ},
  journal={Cell},
  volume={183},
  number={5},
  pages={1249--1263},
  year={2020}
}

@article{behrens2018cognitive,
  title={What is a cognitive map? Organizing knowledge for flexible behavior},
  author={Behrens, Timothy EJ and Muller, Timothy H and Whittington, James CR and Mark, Shirley and Baram, Alon B and Stachenfeld, Kimberly L and Kurth-Nelson, Zeb},
  journal={Neuron},
  volume={100},
  number={2},
  pages={490--509},
  year={2018}
}

@article{shorten2019survey,
  title={A survey on image data augmentation for deep learning},
  author={Shorten, Connor and Khoshgoftaar, Taghi M},
  journal={Journal of Big Data},
  volume={6},
  number={1},
  pages={1--48},
  year={2019}
}

@article{qin2025covt,
  title={Chain-of-Visual-Thought: Teaching VLMs to See and Think Better with Continuous Visual Tokens},
  author={Qin, Yiming and Wei, Bomin and Ge, Jiaxin and Kallidromitis, Konstantinos and Fu, Stephanie and Darrell, Trevor and Wang, Xudong},
  journal={arXiv preprint arXiv:2511.19418},
  year={2025}
}

@article{bigverdi2024perception,
  title={Perception Tokens Enhance Visual Reasoning in Multimodal Language Models},
  author={Bigverdi, Mahtab and Luo, Zelun and Hsieh, Cheng-Yu and Shen, Ethan and Chen, Dongping and Shapiro, Linda G and Krishna, Ranjay},
  journal={arXiv preprint arXiv:2412.03548},
  year={2024}
}

@article{bonnen2024multiview,
  title={Human-level 3D shape perception emerges from multi-view learning},
  author={Bonnen, Tyler and Malik, Jitendra and Kanazawa, Angjoo},
  journal={arXiv preprint},
  year={2024},
  note={Nature, under review}
}

@article{yamins2016using,
  title={Using goal-driven deep learning models to understand sensory cortex},
  author={Yamins, Daniel LK and DiCarlo, James J},
  journal={Nature Neuroscience},
  volume={19},
  number={3},
  pages={356--365},
  year={2016}
}

@article{eichenbaum2000hippocampus,
  title={A cortical--hippocampal system for declarative memory},
  author={Eichenbaum, Howard},
  journal={Nature Reviews Neuroscience},
  volume={1},
  number={1},
  pages={41--50},
  year={2000}
}

@article{oquab2024dinov2,
  title={DINOv2: Learning Robust Visual Features without Supervision},
  author={Oquab, Maxime and Darcet, Timoth{\'e}e and Moutakanni, Th{\'e}o and Vo, Huy and Szafraniec, Marc and Khalidov, Vasil and Fernandez, Pierre and Haziza, Daniel and Massa, Francisco and El-Nouby, Alaaeldin and others},
  journal={Transactions on Machine Learning Research},
  year={2024}
}

@inproceedings{radford2021learning,
  title={Learning transferable visual models from natural language supervision},
  author={Radford, Alec and Kim, Jong Wook and Hallacy, Chris and Ramesh, Aditya and Goh, Gabriel and Agarwal, Sandhini and Sastry, Girish and Askell, Amanda and Mishkin, Pamela and Clark, Jack and others},
  booktitle={International conference on machine learning},
  pages={8748--8763},
  year={2021},
  organization={PmLR}
}

@article{pan2025closing,
  title={Closing the Data-Efficiency Gap Between Autoregressive and Masked Diffusion LLMs},
  author={Pan, Xu and Hahami, Ely and Fan, Jingxuan and Xie, Ziqian and Sompolinsky, Haim},
  journal={arXiv preprint arXiv:2510.09885},
  year={2025}
}

@article{kurth2023replay,
  title={Replay and compositional computation},
  author={Kurth-Nelson, Zeb and Behrens, Timothy and Wayne, Greg and Miller, Kevin and Luettgau, Lennart and Dolan, Ray and Liu, Yunzhe and Schwartenbeck, Philipp},
  journal={Neuron},
  volume={111},
  number={4},
  pages={454--469},
  year={2023},
  publisher={Elsevier}
}

@inproceedings{muttenthaler2023human,
  title={Human alignment of neural network representations},
  author={Muttenthaler, Lukas and Dippel, Jonas and Linhardt, Lorenz and Vandermeulen, Robert A and Kornblith, Simon},
  booktitle={The Eleventh International Conference on Learning Representations},
  year={2023}
}

@article{kumaran2016learning,
  title={What learning systems do intelligent agents need? Complementary learning systems theory updated},
  author={Kumaran, Dharshan and Hassabis, Demis and McClelland, James L},
  journal={Trends in Cognitive Sciences},
  volume={20},
  number={7},
  pages={512--534},
  year={2016}
}

@article{carvalho2025naturalistic,
  title={Naturalistic computational cognitive science: Towards generalizable models and theories that capture the full range of natural behavior},
  author={Carvalho, Wilka and Lampinen, Andrew},
  journal={arXiv preprint arXiv:2502.20349},
  year={2025}
}

@inproceedings{wang2025vggt,
  title={{VGGT}: Visual geometry grounded transformer},
  author={Wang, Jianyuan and Chen, Minghao and Karaev, Nikita and Vedaldi, Andrea and Rupprecht, Christian and Novotny, David},
  booktitle={Proceedings of the Computer Vision and Pattern Recognition Conference},
  pages={5294--5306},
  year={2025}
}

@article{bonnen2021ventral,
  title={When the ventral visual stream is not enough: A deep learning account of medial temporal lobe involvement in perception},
  author={Bonnen, Tyler and Yamins, Daniel LK and Wagner, Anthony D},
  journal={Neuron},
  volume={109},
  number={17},
  pages={2755--2766},
  year={2021},
  publisher={Elsevier}
}

@article{solomon2020semantic,
  title={Semantic search as pattern completion across a concept},
  author={Solomon, Sarah H and Schapiro, Anna C},
  journal={Trends in cognitive sciences},
  volume={24},
  number={2},
  pages={95--98},
  year={2020},
  publisher={Elsevier}
}

@article{stachenfeld2017hippocampus,
  title={The hippocampus as a predictive map},
  author={Stachenfeld, Kimberly L and Botvinick, Matthew M and Gershman, Samuel J},
  journal={Nature neuroscience},
  volume={20},
  number={11},
  pages={1643--1653},
  year={2017},
  publisher={Nature Publishing Group}
}

@article{momennejad2017successor,
  title={The successor representation in human reinforcement learning},
  author={Momennejad, Ida and Russek, Evan M and Cheong, Jin H and Botvinick, Matthew M and Daw, Nathaniel Douglass and Gershman, Samuel J},
  journal={Nature human behaviour},
  volume={1},
  number={9},
  pages={680--692},
  year={2017},
  publisher={Nature Publishing Group UK London}
}

@article{tarder2024brain,
  title={The brain hierarchically represents the past and future during multistep anticipation},
  author={Tarder-Stoll, Hannah and Baldassano, Christopher and Aly, Mariam},
  journal={Nature Communications},
  volume={15},
  number={1},
  pages={9094},
  year={2024},
  publisher={Nature Publishing Group UK London}
}

@article{aly2018flexible,
  title={Flexible weighting of diverse inputs makes hippocampal function malleable},
  author={Aly, Mariam and Turk-Browne, Nicholas B},
  journal={Neuroscience Letters},
  volume={680},
  pages={13--22},
  year={2018},
  publisher={Elsevier}
}

@article{sucholutsky2025getting,
  title={Getting aligned on representational alignment},
  author={Sucholutsky, Ilia and Muttenthaler, Lukas and Weller, Adrian and Peng, Andi and Bobu, Andreea and Kim, Been and Love, Bradley C and Cueva, Christopher J and Grant, Erin and Groen, Iris and others},
  year={2025},
  journal={Transactions on Machine Learning Research}
}

@article{masis2022schema,
  title={Schema representations in distinct brain networks support narrative memory during encoding and retrieval},
  author={Mas{\'\i}s-Obando, Rolando and Norman, Kenneth A and Baldassano, Christopher},
  journal={elife},
  volume={11},
  pages={e70445},
  year={2022},
  publisher={eLife Sciences Publications, Ltd}
}

@article{cutler2019searching,
  title={Searching for semantic knowledge: A vector space semantic analysis of the feature generation task},
  author={Cutler, Rebecca A and Duff, Melissa C and Polyn, Sean M},
  journal={Frontiers in human neuroscience},
  volume={13},
  pages={341},
  year={2019},
  publisher={Frontiers Media SA}
}

@article{foster2006reverse,
  title={Reverse replay of behavioural sequences in hippocampal place cells during the awake state},
  author={Foster, David J and Wilson, Matthew A},
  journal={Nature},
  volume={440},
  number={7084},
  pages={680--683},
  year={2006},
  publisher={Nature Publishing Group UK London}
}

@article{tarder2026hippocampus,
  title={The hippocampus rapidly integrates sequence representations during novel multistep predictions},
  author={Tarder-Stoll, Hannah and Baldassano, Christopher and Aly, Mariam},
    journal = {Philosophical Transactions of the Royal Society B: Biological Sciences},
    volume = {381},
    number = {1954},
    pages = {20250237},
    year = {2026},
    month = {07}
}

@article{schlichting2015learning,
  title={Learning-related representational changes reveal dissociable integration and separation signatures in the hippocampus and prefrontal cortex},
  author={Schlichting, Margaret L and Mumford, Jeanette A and Preston, Alison R},
  journal={Nature communications},
  volume={6},
  number={1},
  pages={8151},
  year={2015},
  publisher={Nature Publishing Group UK London}
}

@book{duhem1954aim,
  title={The Aim and Structure of Physical Theory},
  author={Duhem, Pierre},
  year={1954},
  note={Originally published 1906},
  publisher={Princeton University Press}
}

@book{brindley1960physiology,
  title={Physiology of the Retina and Visual Pathway},
  author={Brindley, Giles S},
  year={1960},
  publisher={Edward Arnold},
  address={London}
}

@article{shohamy2008integrating,
  title={Integrating memories in the human brain: hippocampal-midbrain encoding of overlapping events},
  author={Shohamy, Daphna and Wagner, Anthony D},
  journal={Neuron},
  volume={60},
  number={2},
  pages={378--389},
  year={2008},
  publisher={Elsevier}
}

@article{schwartenbeck2023generative,
  title={Generative replay underlies compositional inference in the hippocampal-prefrontal circuit},
  author={Schwartenbeck, Philipp and Baram, Alon and Liu, Yunzhe and Mark, Shirley and Muller, Timothy and Dolan, Raymond and Botvinick, Matthew and Kurth-Nelson, Zeb and Behrens, Timothy},
  journal={Cell},
  volume={186},
  number={22},
  pages={4885--4897},
  year={2023},
  publisher={Elsevier}
}

@article{he2026human,
  title={Human hippocampal ripples coordinate planning sequences and compositional representations in neocortex},
  author={He, Li and Wang, Xiongfei and Zhang, Jinbo and Xiao, Zhibing and Hu, Xiangyu and Schwartenbeck, Philipp and Bakermans, Jacob and Behrens, Tim and Liu, Yunzhe},
  journal={Nature Neuroscience},
  pages={1--11},
  year={2026},
  publisher={Nature Publishing Group US New York}
}

@article{lewis2011overlapping,
  title={Overlapping memory replay during sleep builds cognitive schemata},
  author={Lewis, Penelope A and Durrant, Simon J},
  journal={Trends in cognitive sciences},
  volume={15},
  number={8},
  pages={343--351},
  year={2011},
  publisher={Elsevier}
}

@article{rasch2007odor,
  title={Odor cues during slow-wave sleep prompt declarative memory consolidation},
  author={Rasch, Bj{\"o}rn and B{\"u}chel, Christian and Gais, Steffen and Born, Jan},
  journal={Science},
  volume={315},
  number={5817},
  pages={1426--1429},
  year={2007}
}

@article{rudoy2009strengthening,
  title={Strengthening individual memories by reactivating them during sleep},
  author={Rudoy, John D and Voss, Joel L and Westerberg, Carmen E and Paller, Ken A},
  journal={Science},
  volume={326},
  number={5956},
  pages={1079--1079},
  year={2009}
}

@article{mattar2018prioritized,
  title={Prioritized memory access explains planning and hippocampal replay},
  author={Mattar, Marcelo G and Daw, Nathaniel D},
  journal={Nature neuroscience},
  volume={21},
  number={11},
  pages={1609--1617},
  year={2018},
  publisher={Nature Publishing Group US New York}
}

@article{guest2021computational,
  title={How computational modeling can force theory building in psychological science},
  author={Guest, Olivia and Martin, Andrea E},
  journal={Perspectives on Psychological Science},
  volume={16},
  number={4},
  pages={789--802},
  year={2021},
  publisher={Sage Publications Sage CA: Los Angeles, CA}
}

@article{kim2026data,
  title={Data-efficient pre-training by scaling synthetic megadocs},
  author={Kim, Konwoo and Kotha, Suhas and Choi, Yejin and Hashimoto, Tatsunori and Haber, Nick and Liang, Percy},
  journal={arXiv preprint arXiv:2603.18534},
  year={2026}
}

\end{document}